%% file: main.tex
\documentclass[preprintnumbers,aps,prd,groupedaddress,twocolumn,floatfix,letterpaper]{revtex4-2}

\pdfoutput=1

\usepackage[T1]{fontenc}
\usepackage[utf8]{inputenc}
\usepackage{amsmath,amssymb,amsfonts,mathtools,bm}
\usepackage[colorlinks=true, citecolor=blue, linkcolor=blue]{hyperref}
\usepackage[final]{graphicx}
\usepackage{xcolor}
\usepackage{booktabs}
\usepackage{relsize}
\usepackage{lineno}
\usepackage{comment}
\usepackage{braket}
\usepackage{dsfont}
\usepackage{float}
\usepackage{hyperref}
\usepackage{needspace}
\usepackage{enumitem}

\usepackage{lineno}

\usepackage{csquotes}

\usepackage{algpseudocode}

\makeatletter
\newcounter{algorithm}
\renewcommand{\thealgorithm}{\arabic{algorithm}}

\newcommand{\ALG@caption}[1]{%
  \refstepcounter{algorithm}%
  \normalsize\noindent\textbf{Algorithm~\thealgorithm.}\ #1\par
  \vspace{2pt}\hrule\vspace{6pt}%
}

\newenvironment{algorithm}[1][t]{%
  \begin{figure}[#1]
  \begin{minipage}{\columnwidth}
  \hrule\vspace{6pt}%
  \renewcommand{\caption}[1]{\ALG@caption{##1}}%
}{%
  \vspace{6pt}\hrule
  \end{minipage}
  \end{figure}
}
\makeatother

\algrenewcommand\algorithmiccomment[1]{\hfill$\triangleright$~#1}

\usepackage{etoolbox}
\makeatletter
\pretocmd{\bibitem}{%
  \ifnum\value{NAT@ctr}=7\relax
    \newpage 
  \fi
}{}{}
\makeatother

\def\Babar{{\mbox{\slshape B\kern-0.1em{\smaller A}\kern-0.1em B\kern-0.1em{\smaller A\kern-0.2em R}}}}

\usepackage{cancel}

\usepackage{soul}

\usepackage{siunitx} 
\usepackage[caption=false]{subfig}

\setstcolor{red}

\begin{document}

\preprint{TUM-EFT 200/25}

\title{Differentiable quantum-trajectory simulation of Lindblad dynamics for QGP transport-coefficient inference}

\author{Lukas Heinrich}
\email{l.heinrich@tum.de}
\affiliation{Technical University of Munich, TUM School of Natural Sciences, Physics Department, James-Franck-Strasse 1, 85748 Garching, Germany}

\author{Tom Magorsch}
\email{tom.magorsch@tum.de}
\affiliation{Technical University of Munich, TUM School of Natural Sciences, Physics Department, James-Franck-Strasse 1, 85748 Garching, Germany}

\begin{abstract}
We study parameter estimation for the transport coefficients of the quark-gluon plasma by differentiating open-quantum-system-based Monte Carlo simulations of quarkonium suppression. The underlying simulator requires solving a Lindblad equation in a large Hilbert space, which makes parameter estimation computationally expensive. We approach the problem using gradient-based optimization. Specifically, we apply the score-function gradient estimator to differentiate through discrete jump sampling in the Monte Carlo wave-function algorithm used to solve the Lindblad equation. The resulting stochastic gradient estimator exhibits sufficiently low variance and can still be estimated in an embarrassingly parallel manner, enabling efficient scaling of the simulations. We implement this gradient estimator in the existing open-source quarkonium suppression code QTraj. To demonstrate its utility for parameter estimation, we infer the two transport coefficients $\hat{\kappa}$ and $\hat{\gamma}$ using gradient-based optimization on synthetic nuclear modification factor data.
\end{abstract}

\maketitle

\input{Introduction}

\input{Theory}

\input{Results}

\input{Conclusion}

\section*{Acknowledgments}
L.H. and T.M. thank Ludwig Burger, Nicole Hartman and Annalena Kofler for useful discussion. T.M. thanks Nora Brambilla and Mara Dauber for useful discussion and comments on the draft.
L.H. and T.M. acknowledge support by the DFG cluster of excellence ORIGINS funded by the Deutsche Forschungsgemeinschaft (DFG) under Germany's Excellence Strategy - EXC-2094-390783311. The authors gratefully acknowledge the Gauss Centre for Supercomputing e.V. (\href{www.gauss-centre.eu}{www.gauss-centre.eu}) for
funding this project by providing computing time on the GCS Supercomputer SuperMUCNG at the Leibniz Supercomputing Centre (\href{www.lrz.de}{www.lrz.de}).

\appendix
\input{Appendix}

\bibliographystyle{apsrev4-1}
\renewcommand*{\bibfont}{\footnotesize}
\bibliography{lit.bib}

\end{document}

%% file: Introduction.tex
\section{Introduction}

Microseconds after the Big Bang, the universe was filled with a hot liquid-like state of matter called the quark-gluon plasma (QGP). This novel phase of matter can be recreated in collider experiments, and its study is one of the central goals of the heavy-ion programs at the LHC and RHIC~\cite{Busza:2018rrf}. 
One key signature to probe the properties of the QGP is quarkonium suppression. While propagating through the QGP, the hot medium dissociates heavy quark-antiquark bound states, leading to a reduced yield in heavy-ion collisions, compared with proton-proton collisions~\cite{Matsui:1986dk}. This suppression is measured by the nuclear modification factor $R_{AA}$. 
A first-principles description of this phenomenon, which has gained traction in recent years, is based on the combination of non-relativistic effective field theories of quantum chromodynamics and the open quantum system paradigm~\cite{Brambilla:2017zei,Brambilla:2022ynh,Akamatsu:2020ypb,Yao:2021lus}. It describes the evolution of quarkonium with a Lindblad equation~\cite{Lindblad:1975ef,Gorini:1975nb}, i.e., a partial differential equation for the time evolution of the density matrix. Making phenomenological predictions, therefore, requires an expensive simulator that solves this equation. Current implementations rely on a Monte Carlo algorithm called the quantum trajectory algorithm~\cite{Molmer:1993ltv,Daley:2014fha}. This algorithm estimates the solution by averaging over independent stochastic trajectories.
Importantly, the simulator depends on specific properties of the QGP characterized by so-called transport coefficients. 
These transport coefficients encode intrinsic properties of the QGP and provide a quantitative handle on its transport behavior, such as viscosity. Comparing the simulator's predictions with measurements of quarkonium suppression thus enables us to infer transport properties of the medium. We therefore face a parameter estimation problem with a simulator that is expensive to evaluate. 

A common black-box approach in such settings is Bayesian optimization~\cite{bayesianopt}, which utilizes successive simulator evaluations to train a surrogate model of the target function and proposes new parameter configurations to evaluate. Bayesian optimization has the advantage that it is usually easy to deploy in a plug-and-play fashion for any black-box simulator, and it provides uncertainty estimates from the surrogate of the objective function. Bayesian optimization is used for transport-coefficient inference in heavy-ion collision phenomenology, primarily in classical or semiclassical transport and effective modeling frameworks \cite{JETSCAPE:2021ehl,Nijs:2020roc,Paquet:2023rfd}. In contrast, the open quantum system approach considered here requires large-scale Lindblad quantum trajectory simulations in a high-dimensional Hilbert space, making black-box inference more costly, especially, with increasing dimensionality of the parameter space. Therefore, parameter inference on open-quantum-system-based quarkonium suppression simulations has not been performed.
An alternative is gradient-based optimization, which can scale more favorably with increasing parameter dimension and can converge with fewer simulator evaluations in many settings. However, it demands the gradient of the simulator output with respect to the parameters, essentially requiring us to \enquote{open up} the black-box simulator. Obtaining this gradient is particularly challenging for a stochastic simulator, which samples a distribution that depends on the parameters in question. 

Significant progress has been made in automatic differentiation (AD) as a tool for computing gradients of complex computations~\cite{Baydin:2015tfa}. While AD, in its original form, applies to deterministic computational graphs, it can be extended to the differentiation of arbitrary stochastic computations using stochastic gradient estimators~\cite{schulman2015gradient,krieken2021storchastic}. Usually, observables from a Monte Carlo simulation can be written as an expectation value $\mathbb{E}_{x\sim p(x|\Theta)}[f(x,\Theta)]$ over some distribution $p(x|\Theta)$, with input parameters $\Theta$. Computing the gradient of this expectation value is a stochastic gradient problem
and is closely related to policy-gradient methods in reinforcement learning. It becomes particularly challenging when $x$ involves discrete random variables, as is often the case in complex Monte Carlo simulations.
In the case of discrete random variables, methods like reparametrization cannot be used; instead, a general-purpose stochastic gradient estimator that is applicable is the score-function gradient estimator, initially introduced as REINFORCE~\cite{Williams:1992mfq,suttonReinforce} in the context of reinforcement learning. Score-function-based differentiation of Monte Carlo expectation values has previously been explored in the high-energy physics context~\cite{Brehmer:2020brs,Nachman:2022jbj,Kagan:2023gxz}.

In this work, we differentiate the open-quantum-system-based quarkonium suppression simulation to enable gradient-based optimization of the QGP transport coefficients. In particular, we 
\begin{enumerate}[topsep=4pt,itemsep=2pt,leftmargin=    
16pt]
    \item differentiate the quantum trajectory algorithm for the solution of the Lindblad equation using the score-function gradient estimator 
    \item show that we obtain a low variance gradient estimate for the survival probabilities of quarkonium states in quarkonium suppression simulations
    \item implement the gradient estimator at scale to perform parameter estimation on synthetic data of the nuclear modification factor using gradient-based optimization
\end{enumerate}

The paper is structured as follows:
In Sec.~\ref{sec:theory}, we provide some background by revisiting quarkonium suppression in the open quantum system framework in Sec.~\ref{sec:suppr} and further reviewing the quantum trajectory algorithm for the Lindblad equation in Sec.~\ref{sec:Lind}. Furthermore, we introduce the score-function gradient estimator in Sec.~\ref {sec:score_grad}.
In Sec.~\ref{sec:results}, we show our results by first applying the score-function gradient estimation to the quantum trajectory algorithm in Sec.~\ref{sec:diff} and applying it to the simulation of quarkonium suppression in Sec.~\ref{sec:diffquarkonium}. In Sec.~\ref{sec:var}, we then compare the variance to a naive finite difference gradient estimation for simple simulations of the Lindblad equation for the quarkonium transport. In Sec.~\ref{sec:opt}, we finally demonstrate an optimization, rediscovering transport coefficient values on a synthetic dataset, and conclude in Sec.~\ref{sec:conc}.

%% file: Theory.tex
\section{Theory}
\label{sec:theory}

We lay out the necessary background by establishing the open quantum system approach to quarkonium suppression and outlining how we obtain phenomenological predictions. We then discuss the quantum trajectories algorithm for solving the Lindblad equation and present the score-function gradient estimator.

\subsection{Quarkonium suppression as an open quantum system}
\label{sec:suppr}

In recent years, major progress has been made in describing the dynamics of heavy-quarkonium in a medium by combining non-relativistic effective field theories with the open quantum system framework~\cite{Akamatsu:2020ypb,Yao:2021lus}. In particular, it is possible to derive an evolution equation for the density matrix $\rho$ of the quarkonium directly from quantum field theory. Using potential non-relativistic QCD Refs.~\cite{Brambilla:2017zei,Brambilla:2022ynh} derived a master equation for the time evolution of bottomonium in the QGP. Here, the density matrix $\rho(\vec{r},\vec{r}^{\,\prime})$ depends on the relative coordinate $\vec{r}$ between the quark-antiquark pair; in the following, we suppress the dependence on $(\vec{r},\vec{r}^{\,\prime})$ for readability. For large temperatures, it is possible to bring the master equation into the Lindblad form~\cite{Lindblad:1975ef,Gorini:1975nb}
\begin{align}
    \begin{split}
        \frac{d\rho(t)}{dt}=-i[&H,\rho(t)]\\
    &+\sum_{n}\left[C_n\rho(t)C_n^{\dagger}-\frac{1}{2}\left\{C_n^{\dagger}C_n,\rho(t)\right\}\right],
        \label{eq:lindblad}
    \end{split}
\end{align}
where $H$ is the Hamiltonian and the $C_n$ are called collapse or jump operators, which encode the effect of the thermal environment. We give their explicit form for the master equation of the quarkonium in Appendix~\ref{app:quarksuppr}.
Both the Hamiltonian and the jump operators depend on dimensionless parameters $\hat\kappa$ and $\hat\gamma$, which are referred to as transport coefficients. The transport coefficients encode the effects of the medium on the quarkonium. They have a field-theoretic definition in terms of correlation functions of the chromo-electric field; however, they are inherently non-perturbative objects, making a first-principles calculation challenging~\cite{Brambilla:2019tpt,Scheihing-Hitschfeld:2023tuz,Brambilla:2025cqy,Brambilla:2025xnw,Brambilla:2024tqg}. We therefore aim to infer the transport coefficients from data using the theory predictions obtained from the Lindblad equation.

To solve the Lindblad equation Eq.~\eqref{eq:lindblad} in three spatial dimensions, we typically project the density matrix onto spherical harmonics in practice. Assuming isotropy, this leads to an evolution equation in a single spatial dimension, the relative radial coordinate $r$, and an angular momentum quantum number $l$. Since the quarkonium can be in a color-singlet or color-octet configuration, the density matrix also depends on a color quantum number $c$.
The jump operators $C_n$ act on the density matrix in position space and induce discrete angular momentum and color transitions. 

From the time evolution of $\rho(t)$, we compute the survival probabilities of different quarkonium states while traversing through the plasma. For each of the states, we can then calculate the nuclear modification factor $R^{i}_{AA}$, which quantifies the suppression of a given state $i$ relative to proton-proton collisions. The nuclear modification factor is measured in heavy-ion collision experiments~\cite{ATLAS:2022exb,ALICE:2020wwx,CMS:2018zza,CMS:2023lfu} and enables a direct test of theory predictions. We provide more detail on how we obtain the nuclear modification factor from the solution of the Lindblad equation $\rho(t)$ in Appendix~\ref{app:pheno}.

\subsection{Lindblad equation and quantum trajectories}
\label{sec:Lind}

After projecting the Lindblad equation~\eqref{eq:lindblad} onto spherical harmonics, we obtain an equation for the density matrix, $\rho$, which now depends on the color $c$, angular momentum $l$, and the radial coordinate $r$. When discretizing the position space coordinate and placing a cutoff on the angular momentum space, we still end up with a very large Hilbert space. As the size of the density matrix scales with the square of the Hilbert space dimension, solving the Lindblad equation is a challenging problem. An efficient and scalable approach is the unraveling of the dynamics in terms of stochastic trajectories, also known as the Monte Carlo wave function or quantum trajectories method~\cite{Molmer:1993ltv,Daley:2014fha}. The idea is, instead of evolving the full density matrix $\rho$, to calculate a stochastic evolution of pure states $\ket{\psi_i}$, which, when averaging over many trajectories, restores the full dynamics of the density matrix.

In Algorithm~\ref{alg:waiting_time_mcwf}, we give the quantum trajectories algorithm for sampling the evolution of individual trajectories. The algorithm takes a Hamiltonian $H$ and $N_C$ jump operators $C_n$ and returns the time evolution of trajectories $\ket{\psi(t)}$ given an initial state $\ket{\psi_0}=\ket{\psi(t_0)}$. The algorithm is presented in the \textit{waiting-time approach}, which evolves the wave function with the effective Hamiltonian $H_\text{eff}$ over time steps $\Delta t$. Since the effective Hamiltonian is non-Hermitian, the time evolution will reduce the norm of the wave function so that the norm $\braket{\psi|\psi}$ keeps decreasing. A quantum jump is triggered once the norm drops below a random threshold $r$, which is drawn from a uniform distribution. When a quantum jump occurs, a random jump operator is drawn and applied to the state, after which the wave function is finally normalized, before the evolution continues. At each time step, the algorithm decides whether the state undergoes no-jump evolution under $H_\text{eff}$, or a quantum jump occurs and, if so, which jump operator $C_n$ is applied to the state. This procedure is equivalent to sampling a discrete random variable from a categorical distribution with $N_C+1$ categories.

\begin{algorithm}[t]
\caption{\normalsize {Waiting-time quantum trajectories algorithm}}
\label{alg:waiting_time_mcwf}
\begin{algorithmic}[1]
\Require $H$, $\{C_n\}_{n=1}^{N_C}$, $\Delta t$, $\ket{\psi_0}$, $t_0,t_{\max}$
\Ensure One trajectory $\ket{\psi(t)}$ for $t\in[t_0,t_\text{max}]$

\State $t \gets t_0$, \quad $\ket{\psi} \gets \ket{\psi_0}$
\State $r \sim \mathrm{Uniform}(0,1)$
\State $H_{\mathrm{eff}} \gets H - \frac{i}{2}\sum_{n=1}^{N_C} C_n^\dagger C_n$

\While{$t < t_{\max}$}
  \State $\ket{\psi} \gets \exp(-i H_{\mathrm{eff}}\Delta t)\ket{\psi}$
  \If{$\braket{\psi|\psi} < r$} \Comment{quantum jump}
    \State $w_n \gets \bra{\psi} C_n^\dagger C_n \ket{\psi}$ \ \textbf{for all} $n\in\{1,\dots,N_C\}$
    \State $p_n \gets w_n \big/ \sum_m w_m$
    \State Sample $n \sim \mathrm{Categorical}(p_1,\ldots,p_{N_C})$
    \State $\ket{\psi} \gets C_n\ket{\psi}/\|C_n\ket{\psi}\|$
    \State $r \sim \mathrm{Uniform}(0,1)$
  \EndIf
  \State $t \gets t + \Delta t$
  \State Record $\ket{\psi}$ at time $t$
\EndWhile
\State \Return $\ket{\psi(t)}$ at $t\in[t_0,t_\text{max}]$
\end{algorithmic}
\end{algorithm}

It can be shown that averaging over many of these stochastic pure state trajectories recovers the correct evolution of the density matrix~\cite{Daley:2014fha}
\begin{equation}
\rho(t) = \mathbb{E}\left[\ket{\Psi(t)}\bra{\Psi(t)}\right], 
\end{equation}
where we define the normalized trajectory
\begin{equation}
    \ket{\Psi(t)}=\frac{\ket{\psi(t)}}{\sqrt{\braket{\psi(t)|\psi(t)}}},
    \label{eq:psinorm}
\end{equation}
since in the waiting-time approach the sampled trajectories $\ket{\psi(t)}$ are not normalized.
Similarly, an observable $O$ can be calculated by averaging over the observable for the different trajectories
\begin{equation}
    \braket{O(t)} = \text{Tr}(O\rho(t)) = \mathbb{E}\left[\braket{\Psi(t)|O|\Psi(t)}\right].
    \label{eq:expect}
\end{equation}
In practice, instead of recording the full trajectories to disc, it is usually sufficient to only save specific observables per-trajectory at specific times, considerably reducing the required memory.  
This stochastic algorithm is particularly convenient because it is embarrassingly parallel among the different trajectories, allowing for efficient large-scale computations.
The quantum trajectories algorithm was implemented for the non-relativistic evolution of quarkonium in medium in the open-source code QTraj~\cite{Omar:2021kra}.

\subsection{Score-function gradient estimator}
\label{sec:score_grad}

To calculate gradients of the quantum trajectories simulation results, we have to differentiate Eq.~\eqref{eq:expect}. This setting corresponds to the general problem of stochastic gradient estimation, where, given an expectation 
\begin{equation}
    \mathbb{E}_{p(x|\Theta)}\left[f(x,\Theta)\right],
\end{equation}
we want to estimate
\begin{equation}
    \frac{d}{d\Theta}\mathbb{E}_{p(x|\Theta)}\left[f(x,\Theta)\right].
    \label{eq:gradient}
\end{equation}
For simplicity, we consider $\Theta$ to be a scalar; the generalization to multiple parameters is straightforward. 
A naive finite-difference estimation of the form
\begin{align}
    \begin{split}
    \frac{d}{d\Theta}&\mathbb{E}_{p(x|\Theta)}\left[f(x,\Theta)\right]\\&=\frac{\mathbb{E}_{p(x|\Theta+\epsilon)}\left[f(x,\Theta+\epsilon)\right]-\mathbb{E}_{p(x|\Theta-\epsilon)}\left[f(x,\Theta-\epsilon)\right]}{2\epsilon},
    \end{split}
    \label{eq:finitediff}
\end{align}
is both expensive to evaluate and suffers from a large variance, as the two expectations are sampled independently. Furthermore, it suffers from a bias-variance tradeoff, as a large $\epsilon$ will bias the gradient, while a small $\epsilon$ leads to a large variance.

In some cases, it is possible to reparametrize this gradient~\cite{kingma2022autoencodingvariationalbayes} by finding a transformation $g(\epsilon,\Theta)$ so that $x=g(\epsilon,\Theta)$, where $\epsilon\sim p(\epsilon)$ is an independent random variable. 
Given such a reparametrization, we can write the gradient as
\begin{align}
\begin{split}
    \frac{d}{d\Theta}\mathbb{E}_{p(\epsilon)}&[ f(g(\epsilon,\Theta),\Theta)]\\ &=\mathbb{E}_{p(\epsilon)}\left[\frac{\partial f(x,\Theta)}{\partial x}\frac{\partial g(\epsilon,\Theta)}{\partial \Theta} + \frac{\partial f(x,\Theta)}{\partial \Theta}\right].
    \end{split}
\end{align}
However, for this to be possible, $f(x,\Theta)$ needs to be differentiable in $x$ and $g(\epsilon,\Theta)$ needs to be differentiable in $\Theta$. Therefore, often such a reparametrization does not exist, especially when $x$ takes discrete values. A more general scheme to estimate the gradient Eq.~\eqref{eq:gradient} is based on the log-derivative trick, originally introduced as REINFORCE gradient estimator~\cite{Williams:1992mfq,suttonReinforce,mohamed2020monte}. This estimator, also known as score-function gradient estimator, uses the score function to rewrite the gradient of the expectation as another expectation over the same distribution
\begin{align}
\begin{split}
    \frac{d}{d\Theta}\mathbb{E}_{p(x|\Theta)}\left[f(x,\Theta)\right]=\int dx\, \frac{d}{d\Theta}\left[p(x|\Theta)f(x,\Theta)\right]\\
    =\int dx \,p(x|\Theta)\left[\frac{d\log p(x|\Theta)}{d\Theta}f(x,\Theta)+\frac{d}{d\Theta}f(x,\Theta)\right]\\
    =\mathbb{E}_{p(x|\Theta)}\left[\frac{d\log p(x|\Theta)}{d\Theta}f(x,\Theta)+\frac{d}{d\Theta}f(x,\Theta)\right].
\end{split}    
    \label{eq:scoreestimate}
\end{align}
The estimator consists of two terms. The first term, which we will refer to as the \textit{score-function term}, accounts for changes in the sampling distribution. The second term, which we will refer to as the \textit{pathwise derivative}, accounts for the direct dependence of the estimator on the parameters. The combination of the score-function and pathwise-derivative terms yields an unbiased gradient estimator which is expected to exhibit a much smaller variance than the finite-difference estimator Eq.~\eqref{eq:finitediff}, as we estimate the gradient on a single set of samples.

To further reduce the variance, it is possible to subtract a \textit{control variate} $c(x,\Theta)$ from the estimator~\cite{greensmith2004variance}. As long as $\mathbb{E}_{p(x|\Theta)}[c(x,\Theta)]=0$ the estimator will remain unbiased. Since for any estimator $g$
\begin{equation}
    \text{Var}(g-c) = \text{Var}(g)+\text{Var}(c)-2\text{Cov}(g,c),
\end{equation}
this subtraction can reduce the variance if the control variate $c$ is correlated with the original estimator $g$.
Finding a particularly good control variate can be challenging. A common default choice is 
\begin{equation}
    c(x,\Theta) = b(\Theta)\frac{d\log p(x|\Theta)}{d\Theta},
    \label{eq:cv}
\end{equation}
with some $b(\Theta)$ which does not depend on $x$. Since the score function has vanishing expectation, the estimator remains unbiased. A common choice for $b(\Theta)$ is to use the expectation of the estimator
\begin{equation}
    b(\Theta) = \mathbb{E}_{p(x|\Theta)}[f(x,\Theta)],
\end{equation}
which can be estimated using the same samples as the original estimator.
This choice is also referred to as the \textit{mean baseline}. As the control variate Eq.~\eqref{eq:cv} is correlated with the score-function estimator, in practice this usually leads to a modest to large variance reduction. In the following, we will use the mean baseline in all experiments.

Using the gradient estimator Eq.~\eqref{eq:scoreestimate}, it is possible to differentiate the output of any stochastic algorithm~\cite{schulman2015gradient} by considering the computation as a graph, consisting of stochastic and deterministic nodes. The stochastic nodes sample random variables from a probability distribution, while the deterministic nodes perform computations of arbitrary functions. Such a graph can be differentiated by applying the score-function estimator to the stochastic nodes of the computational graph. If the computational graph also contains paths from the observable $f$ to the parameters $\Theta$, which do not pass through stochastic nodes, they are accounted for by the pathwise-derivative term of the estimator. In the following, we will use this differentiation scheme to calculate gradients of observables using the quantum trajectories algorithm.

%% file: Results.tex
\section{Results}
\label{sec:results}

We apply the previously introduced score-function gradient estimator to the quantum trajectories algorithm to obtain gradients of trajectory-averaged observables. We then demonstrate the gradient estimator on the simulation of quarkonium suppression by assessing its variance and performing an optimization to determine transport coefficients from nuclear modification factor data.

\subsection{Differentiable quantum trajectories}
\label{sec:diff}

\begin{figure*}[t!]  
    \centering
    \includegraphics[width=0.75\linewidth]{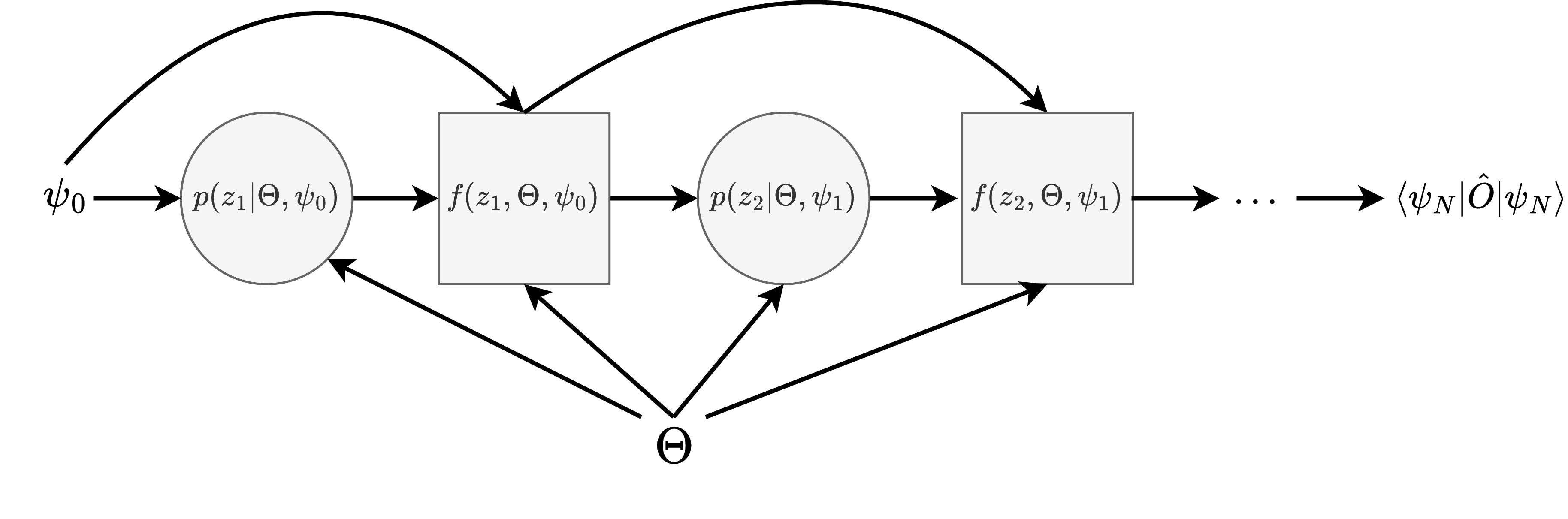}
    \caption{The computational graph for the quantum trajectory algorithm. Squares depict deterministic nodes, which compute a function $f$ on their inputs, while circles represent stochastic nodes, which sample a random variable from a distribution $p$, which can depend on input parameters as well. Here $\Theta$ is a set of parameters, and we assume that the Hamiltonian $H$ and the jump operator $C$ both depend on these parameters. For more details on the definition of $p$ and $f$, see the text.}
    \label{fig:QtrajGraph}
\end{figure*}

The computational graph of the quantum trajectories algorithm is shown in Fig.~\ref{fig:QtrajGraph}. For simplicity, we restrict ourselves to a single jump operator $C$ for now.
Here $\Theta$ denotes the parameters we want to differentiate with respect to. At every timestep, we compute the non-unitary time evolution $U\ket{\psi(t)}$ and then perform a discrete decision whether we perform a jump by applying the jump operator, or not. Here $U \equiv \exp(-iH_{\mathrm{eff}}\Delta t)$ is the non-unitary time evolution operator with $H_\text{eff}=H-\frac{i}{2}C^\dagger C$. In practice, $U$ is applied via a time-stepping scheme. The algorithm thus consists of consecutive stochastic and deterministic nodes, which are represented by circles and squares, respectively. Since in the waiting-time algorithm, outlined in Sec.~\ref{sec:Lind}, the jump probability is given by the decay of the norm per time step, the stochastic node at step $n$ draws a discrete random number $z_n$ from a Bernoulli distribution given by 
\begin{equation}
p(z_n|\Theta,\psi_{n}) =\begin{cases}1-\delta p_n , &z_n=0\\\delta p_n,&z_n = 1\end{cases},
\label{eq:pn}
\end{equation}
where $\psi_n$ is the wave function after $n$ steps and $\delta p_n$ denotes the reduction of the norm through one step of the non-Hermitian evolution. 
This reduction of the norm gives the jump probability per time step, and can be calculated from the time evolution of $\ket{\psi_n}$ as
\begin{equation}
    \delta p_n = 1-\frac{\braket{\psi_n|U^\dagger U|\psi_n}}{\braket{\psi_n|\psi_n}}.
    \label{eq:jump_prob}
\end{equation}
The cases $z_n=1$ and $z_n=0$ represent the jump and no-jump case repectively.
The extension to multiple jump operators replaces Eq.~\eqref{eq:pn} with a categorical distribution, with an additional category per jump operator.
Given the decision $z_n$, the new wave function is computed as
\begin{align}
    \begin{split}
\ket{\psi_{n+1}}=&f(z_n,\Theta,\psi_{n})\\ =& (1-z_n)U\ket{\psi_n}+ z_n\frac{CU\ket{\psi_n}}{||CU\ket{\psi_n}||},
    \end{split}
    \label{eq:f}
\end{align}
which applies the jump operator $C$ only if $z_n=1$.
In practice, it is not necessary to draw a new random number each step; instead, one can draw a single random number once and evolve the state with $U$ until the squared norm drops below this random number as described for the waiting-time approach in Sec.~\ref{sec:Lind}. This survival threshold prescription is statistically equivalent to the sequential Bernoulli draws at each time step.
To apply the score-function gradient estimation discussed in Sec.~\ref{sec:score_grad}, we notice that the averaging of the trajectories is an expectation value over the product of all Bernoulli decisions of the form~\eqref{eq:pn}, so that for any observable $O$ at some fixed time $t$ corresponding to $N$ steps, we have
\begin{equation}
    \braket{O(t)} = \mathbb{E}_{p(z|\Theta)}\left[\braket{\Psi_N|O|\Psi_N}\right],
\end{equation}
where the distribution for all discrete decisions $z=(z_0,\dots,z_N)$ is given by
\begin{equation}
    p(z|\Theta)=\prod_n p(z_n|\Theta,\psi_n),
\end{equation}
and the observable is again evaluated on the normalized state~\eqref{eq:psinorm}.
The score-function gradient estimator~\eqref{eq:scoreestimate} therefore becomes
\begin{align}
\begin{split}
\nabla_\Theta\braket{O(t)}=\mathbb{E}_{p(z|\Theta)}\bigg[&\big(\braket{\Psi_N|O|\Psi_N}-b\big)\\&\times\sum_n \nabla_\Theta\log\big(p(z_n|\Theta,\psi_n)\big)\\  &+\nabla_\Theta\braket{\Psi_N|O|\Psi_N}\bigg],
    \label{eq:qtraj_scoregrad}
\end{split}    
\end{align}
where the $b$ denotes the baseline, chosen as the mean $b=\mathbb{E}[\braket{\Psi_N|O|\Psi_N}]$.
The first term of this estimator is the score-function term, which accounts for the change in jump probabilities resulting from changes in the parameters $\Theta$. It can be computed conveniently by keeping a single running variable and adding the score of the event at every step to it. 
To obtain the score, we take the derivative of Eq.~\eqref{eq:jump_prob} by tracking the time evolution of the gradient $\nabla_\Theta \ket{\psi_n}$, which is obtained by differentiating Eq.~\eqref{eq:f} and can be computed directly or via automatic differentiation from the parameter dependence of the Hamiltonian and jump operator. 
The second term is the pathwise term, and it reflects the fact that in the computational graph Fig.~\ref{fig:QtrajGraph}, there are direct paths from the observable $\braket{O}$ to the parameters $\Theta$. In practice, this contribution can be calculated directly from $\nabla_\Theta\ket{\psi_N}$. Eq.~\eqref{eq:qtraj_scoregrad} provides an unbiased estimator for the gradient of any observable. In particular, it can be implemented with a minimal intervention in the algorithm, thus still enabling the efficient computation by sampling independent trajectories in an embarrassingly parallel manner.

\subsection{Differentiating quarkonium suppression simulations}
\label{sec:diffquarkonium}

We apply the differentiable quantum-trajectories formulation of Sec.~\ref{sec:diff} to the full quarkonium suppression simulator QTraj~\cite{Omar:2021kra}. To generalize the differentiation scheme to the physical simulation, we must consider the different jump operators. After projecting the Lindblad equation for the evolution of the quarkonium Eq.~\eqref{eq:lindblad} on spherical harmonics, we end up with six jump operators $C_m$ with $m=1,...,6$. The jump case $z_n=1$ in Eq.~\eqref{eq:pn} is then split into six cases corresponding to the different jump operators $z_n=1,...,6$. Each jump has the probability 
\begin{equation}
    \delta p^m_n = \delta p_n \frac{\braket{\psi_n|U^\dagger C^\dagger_mC_m U|\psi_n}}{\sum_m \braket{\psi_n|U^\dagger C^\dagger_mC_m U|\psi_n}},
\end{equation}
leading to a categorical distribution in Eq.~\eqref{eq:pn}. The application of the score-function estimator to categorical distribution is identical to the Bernoulli case.
The scores of the probabilities can be determined as in the one-jump case. We apply the estimator with respect to the transport coefficients $\Theta=(\hat\kappa,\hat\gamma)$ to the projection operators $O = \ket{nl}\bra{nl}$, where $\ket{nl}$ are the $n$th eigenstate of the vacuum hamiltonian with angular momentum $l$. The necessary scores and derivatives of the time evolution can be obtained directly from the dependence of the Lindblad equation on the transport coefficients, given in Appendix~\ref{app:quarksuppr}.
This way, we obtain the gradient of the survival probabilities of $S$- and $P$-wave quarkonium states $\nabla_\Theta \braket{nl|\rho(t)|nl}$. As the experimental measurements only depend on the survival probabilities at the end of the evolution through the QGP, we only have to collect gradients at $t=t_\text{max}$. 
In practice, we can save the result for the sum of the scores and the pathwise derivative term to disk, enabling the calculation of Eq.~\eqref{eq:qtraj_scoregrad}. This enables the same parallelization, which permits the QTraj code to scale to large Hilbert space sizes. Since the postprocessing to obtain the nuclear modification factor $R^i_{AA}$ described in Appendix~\ref{app:pheno} is a linear function of the $S$-wave and $P$-wave survival probabilities, propagating the gradient to obtain $\nabla_\Theta R^i_{AA}$ is straightforward.

\subsection{Gradient estimator variance}
\label{sec:var}

\begin{figure*}[t!]  
  \centering
  \includegraphics[width=0.45\textwidth]{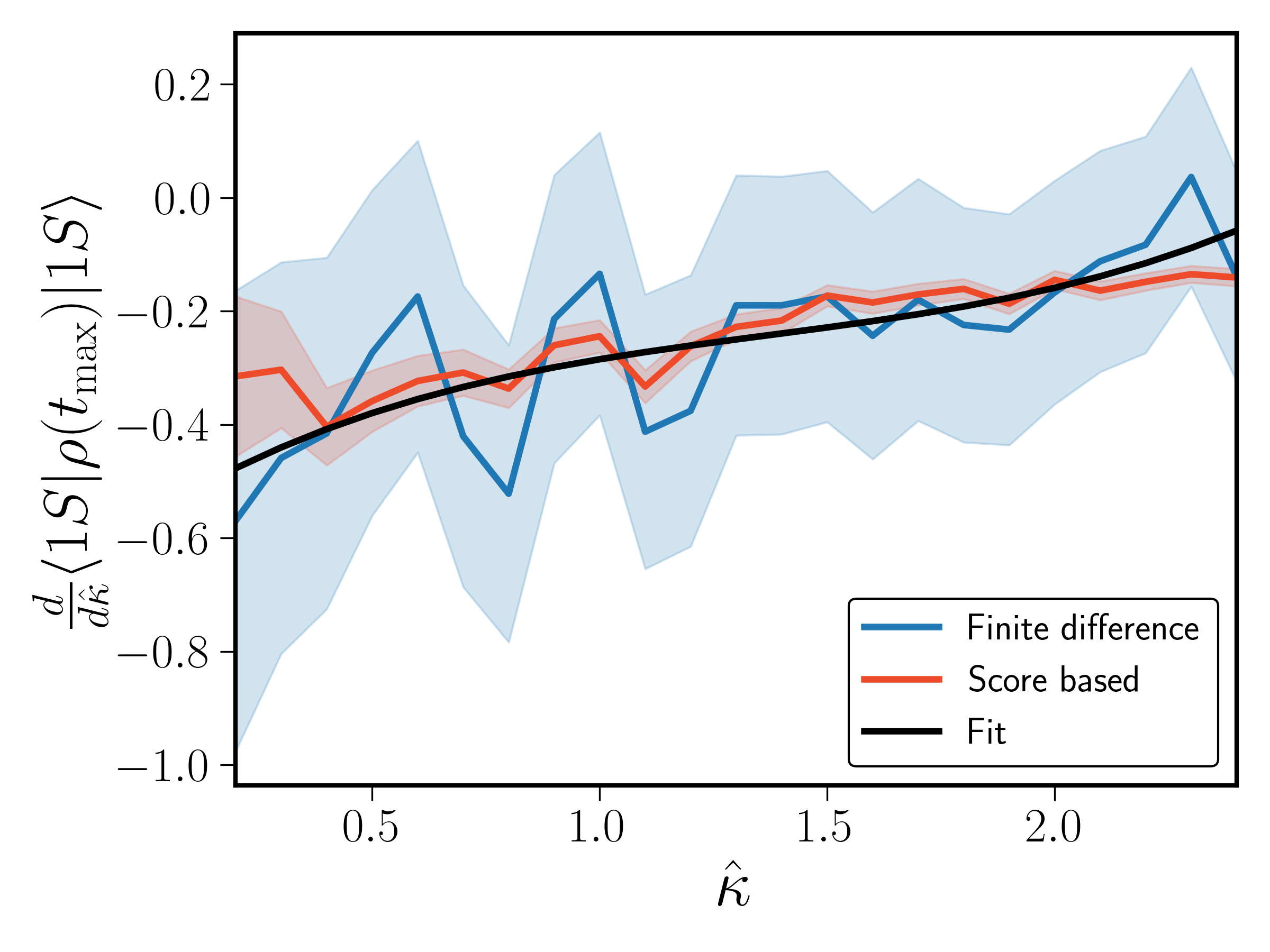}
  \includegraphics[width=0.45\textwidth]{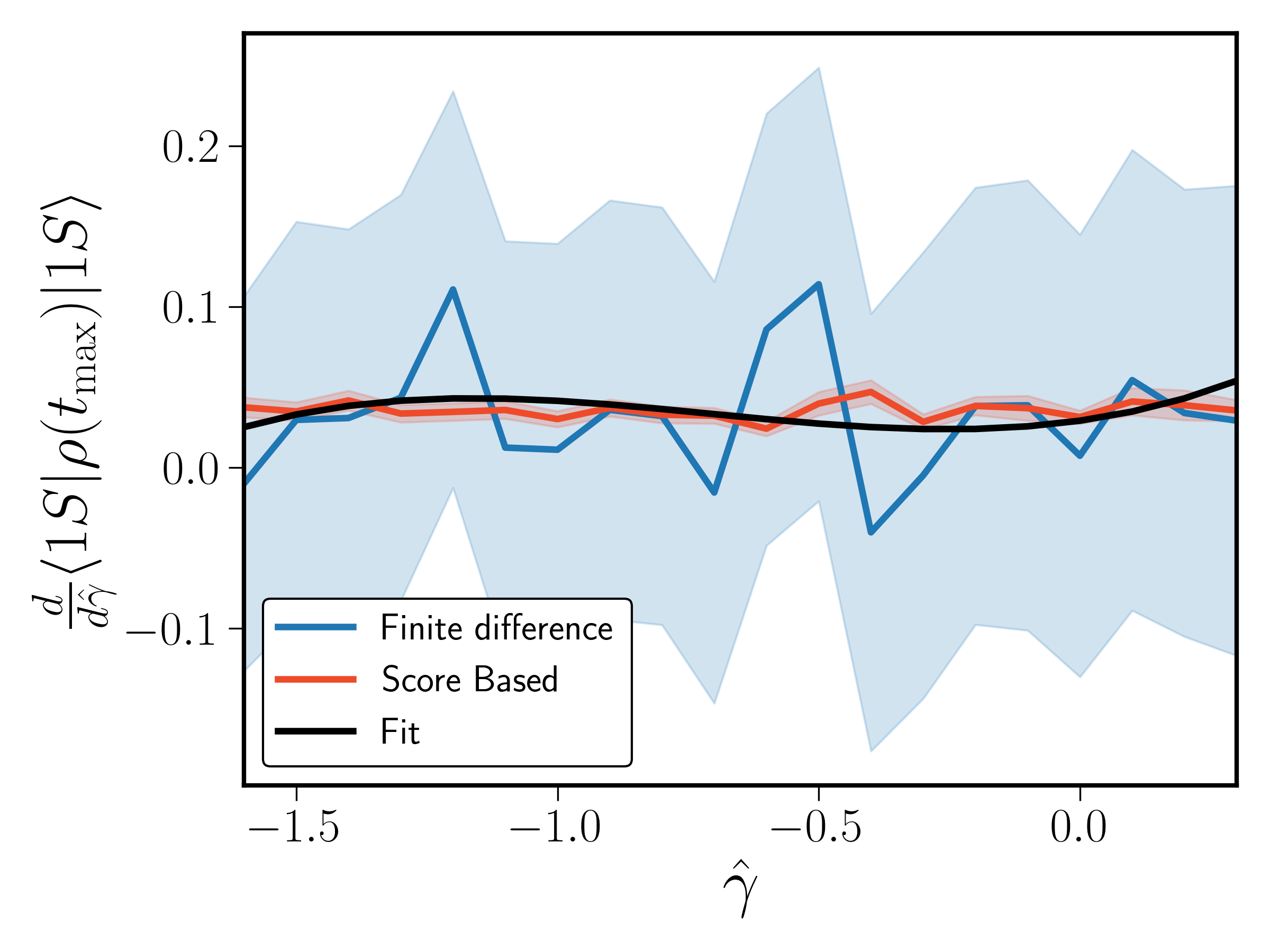}
  \caption{Comparison of gradient estimators for the $1S$ overlap at the end of the evolution $\braket{1S|\rho(t_\text{max})|1S}$. The underlying simulations are performed for a Bjorken temperature evolution, see text. We compare the score-function gradient estimator (red) against a central finite-difference estimation of the gradient (blue). The black line represents the derivative of a polynomial fit, serving as a smoothed estimate of the ground truth. \textit{Left:} The derivative with respect to $\hat{\kappa}$ for fixed $\hat{\gamma}=0$. \textit{Right:} The derivative with respect to $\hat{\gamma}$ for fixed $\hat{\kappa}=4$.}
  \label{fig:variance}
\end{figure*}

We investigate the variance properties of the gradient estimator Eq.~\eqref{eq:qtraj_scoregrad}, by simulating the survival probabilities for different values of the transport coefficients. To conserve computational resources, we simulate the evolution of the quarkonium in a plasma with a Bjorken temperature evolution~\cite{Bjorken:1982qr}, given by 
\begin{equation}
    T(t) = T_0\left(\frac{t_0}{t}\right)^{v^2_s},
\end{equation}
where we choose $T_0 = \SI{450}{\MeV}$, $t_0=\SI{0.6}{\femto\meter}$ and $v^2_s=1/3$.
We perform two sets of simulations: one with $\hat\gamma=0$ fixed and $\hat\kappa\in[0.1,2.5]$, and one with $\hat\kappa=4$ fixed and $\hat\gamma\in[-1.7,0.3]$. In both cases, we perform simulations on a uniform grid of transport coefficient values with a spacing of $0.1$.
We simulate a volume of $L=\SI{40}{\GeV\tothe{-1}}$ and discretize the radial part of the wave function on $N=2048$ points. For more details on the computational setup, we refer the reader to Refs.~\cite{Omar:2021kra,Brambilla:2022ynh}. For each transport coefficient configuration, we sample a total of $\num{10000}$ trajectories. We then estimate the derivative of the $1S$ survival probability at the end of the evolution $\braket{1S|\rho(t_\text{max})|1S}$ with respect to one transport coefficient, once with the score-function estimator Eq.~\eqref{eq:qtraj_scoregrad}, and once using a central finite-difference derivative Eq.~\eqref{eq:finitediff}. For the finite-difference-based estimation, the chosen transport coefficient grids lead to $\epsilon=0.1$. Furthermore, we perform a higher-order polynomial fit to the $1S$ survival probability as a function of the transport coefficient. By taking the derivative of the polynomial, we obtain a smoothed-out estimate for the true gradient. In Fig.~\ref{fig:variance}, in the left and right panels, we show the results for the derivative with respect to $\hat\kappa$ and $\hat\gamma$, respectively. The shaded bands indicate the $\pm 1$ standard error of the mean, estimated from the sample variance across trajectories.
We observe that both gradient estimators, on average, agree well with the polynomial fit. However, the score-function estimator displays a significantly smaller standard error than the finite-difference one, due to its favorable variance properties. This reduced variance eventually makes it possible to apply gradient-based minimizaion of a loss function.

\subsection{Optimization results}
\label{sec:opt}

To demonstrate the ability to perform parameter inference via gradient descent, we draw a synthetic dataset for the nuclear modification factor $R_{AA}$ given the parameters $\hat{\kappa}=4$ and $\hat{\gamma}=0$. We then aim to rediscover these parameters using gradient descent, utilizing the score-function gradient estimator. For the synthetic data, we simulate the evolution of bottomonium in the QGP in lead-lead collisions at the LHC by sampling event-by-event temperature evolutions $T(t)$ for a given number of participating nucleons from anisotropic hydrodynamics simulations~\cite{Alqahtani:2020paa}.
The number of participating nucleons $N_\text{part}$ quantifies how much the nuclei overlap when colliding, and the nuclear modification factor is reported as a function of $N_\text{part}$.
We then solve the Lindblad equation~\eqref{eq:lindblad} for the different temperature profiles using the QTraj code~\cite{Omar:2021kra}. In total, we simulate about $7\,$M trajectories, distributed across $12$ anchor $N_\text{part}$ values and obtain the nuclear modification factor $R_{AA}(N_\text{part})$ and its associated statistical uncertainty $\sigma_{R_{AA}}(N_\text{part})$. Between these $N_\text{part}$, we linearly interpolate the nuclear modification factor and its uncertainty. To obtain a synthetic dataset, we divide the interval of the number of participating nucleons $N_\text{part}$ into equidistant bins and draw a random value of $N_\text{part}$ in each of these bins, leading to a set of randomly drawn values $N^{(k)}_\text{part}$. We then draw the synthetic data for the nuclear modification factor at the $N^{(k)}_\text{part}$ from a normal distribution as 
\begin{equation}
R^{(k)}_{AA}\sim\mathcal{N}\left(R_{AA}(N^{(k)}_\text{part}),\sigma^2_{R_{AA}}(N^{(k)}_\text{part})\right).     
\end{equation}
We repeat this procedure for the $\Upsilon(1S)$, $\Upsilon(2S)$ and $\Upsilon(3S)$ states.
\begin{figure*}[t!]
    \centering    \includegraphics[width=0.7\linewidth]{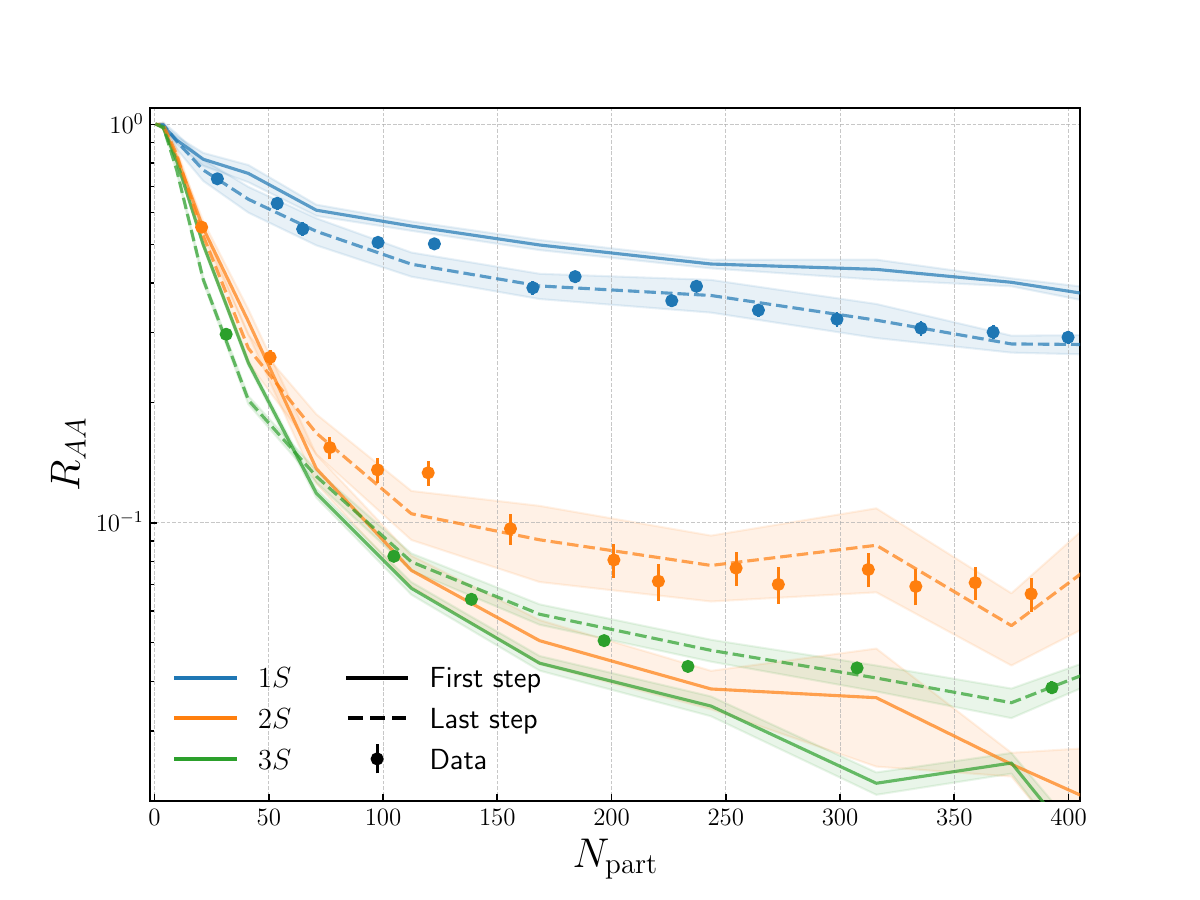}
    \caption{Results for the nuclear modification factor plotted against the number of participating nucleons. The circular data points show the synthetic dataset drawn from a simulation with $\hat{\kappa}=4$ and $\hat{\gamma}=0$. The solid lines represent the simulation results, at the initial parameters $\hat{\kappa}=2$ and $\hat{\gamma}=-1.5$, while the dashed lines show the simulation results at the last step of the optimization. The shaded bands indicate the statistical uncertainty of the simulations.}
    \label{fig:synth_data}
\end{figure*}
This dataset is shown in Fig.~\ref{fig:synth_data} and will serve as the ground truth data generated at the transport coefficient values $\hat{\kappa}=4$ and $\hat\gamma=0$, which we aim to rediscover.

We perform the optimization from the initial parameters $\hat{\kappa}=2,\hat{\gamma}=-1.5$. At each step, we perform a simulation with approximately $1.5\,$M trajectories. To account for the orders of magnitude spanned by $R_{AA}$, we minimize the weighted squared log-ratio
\begin{equation}
    \mathcal{L} = \sum_{i,n} w_{i,n} \ln^2\left(\frac{y_{i,n}}{f_{i,n}}\right),
    \label{eq:loss}
\end{equation}
where the $y_{i,n}$ denote the $i$th datapoint for the nuclear modification factor of the $\Upsilon(nS)$ from the synthetic dataset. The $f_{i,n}$ denotes the corresponding simulator prediction for the nuclear modification factor at the same $N_\text{part}$ for the same state. The $w_{i,n}$ weights each point by its respective uncertainty. We choose to normalize the weights per state
\begin{equation}
    w_{i,n} = \frac{\tilde{w}_{i,n}}{\sum_i\tilde w_{i,n}},
\end{equation}
to prevent one state from dominating the optimization. For $\tilde w_{i,n}$ we use
\begin{equation}
    \tilde w_{i,n}=\frac{1}{\sigma^2_{y_{i,n}}/y^2_{i,n} + \sigma^2_{f_{i,n}}/f^2_ {i,n}},
\end{equation}
where the $\sigma$ denotes the standard deviation of the synthetic data and the simulator prediction, respectively.

Using the gradient estimator for the simulation results of the nuclear modification factor for each state $\nabla_\Theta f_{i,n}$, we compute the gradient $\nabla_\Theta\mathcal{L}$, which in turn is used to optimize $\mathcal{L}$.
We use AMSGrad~\cite{amsgrad} as optimizer with a learning rate of $\eta=0.2$ for fast convergence and momentum parameters $\beta_1 = 0.7$ and $\beta_2=0.99$. 

To make the gradient-descent-based inference feasible at a million-trajectory scale, we run the optimization in an automated HPC workflow by submitting a set of jobs using SLURM, which perform the embarrassingly parallel sampling of trajectories. A Python script then performs the analysis upon completion of all simulation jobs. After post-processing and averaging the trajectories, this script reads in the current state of the optimizer, computes the gradient of the loss function, determines the next transport coefficient configuration, and then specifies the next set of jobs to submit.
\begin{figure*}[t!]
    \centering
    \includegraphics[width=0.7\linewidth]{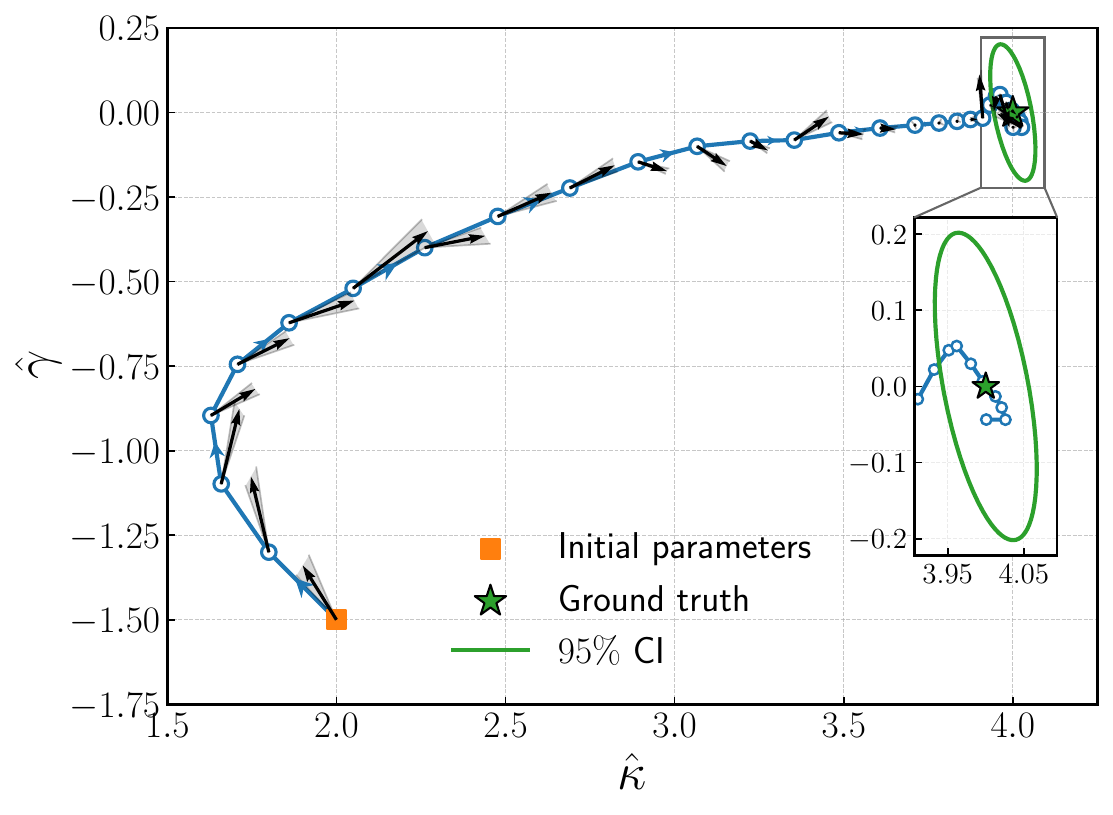}
    \caption{The trajectory of the optimizer in the transport coefficient plane. We perform the optimization using AMSGrad on the synthetic data for the nuclear modification factor, using the loss function~\eqref{eq:loss}. The orange square indicates the initial parameters, open blue circles show the parameter configuration after each step. Black arrows show the estimated gradient of the loss, with the gray cone indicating its statistical uncertainty. Deviations between the gradient and the actual performed step are due to the optimizer. The green star denotes the parameter configuration from which the synthetic data was drawn. The green ellipse around it indicates the $95\%$ confidence interval.}
    \label{fig:opt}
\end{figure*}
The trajectory of the optimization in the $(\hat{\kappa},\hat{\gamma})$ plane is shown in Fig.~\ref{fig:opt}. The orange square indicates the initial parameters, while the optimization trajectory is shown in blue, with each blue circle representing a single step of the optimizer. The black arrows indicate the per-step estimate of the gradient, with the gray cone representing its uncertainty, which we obtain from the variance of the gradient estimation. Deviations between the raw gradient and the blue optimization step are due to the optimizer. The green star displays the ground truth parameters.
To estimate the uncertainty on these ground truth transport coefficient values, originating from the statistical nature of the synthetic dataset, we calculate the $95\%$ confidence interval around this point by estimating the Hessian of the loss $\mathcal{L}$ at the ground truth point. We estimate the Hessian using score-function gradient estimation as well. We indicate the $95\%$ CI by the green ellipse. The optimizer converges in the $95\%$ CI of the ground truth parameters within $\sim 25$ steps. The momentum helps stabilize the optimization against noise, leading to a smooth and controlled path, despite the number of sampled trajectories per step  being low enough to lead to sizable per-step noise. We especially emphasize that the low variance of the score-function gradient estimator enables us to perform optimization with a low number of samples per step. This is especially useful compared to black-box optimization strategies, which compare loss values of independent Monte Carlo simulations at different parameter points, requiring a large number of samples for accurate comparisons. We show the simulation results for the nuclear modification factor for the first and last step of the optimization in Fig.~\ref{fig:synth_data}. The nuclear modification factor results from the last optimization step agree well with the synthetic data, demonstrating the ability of the end-to-end gradient-based optimization to fit the transport coefficients to the nuclear modification factor data.

%% file: Conclusion.tex
\section{Conclusions}
\label{sec:conc}

In this work, we applied the score-function gradient estimator to the quantum trajectories algorithm to differentiate the simulation of Lindblad equations.
We employed this differentiable quantum trajectories algorithm for the open quantum system simulation of quarkonium suppression in heavy-ion collisions at the LHC. 
We obtained a low-variance estimator for the gradients of survival probabilities, which still allows for an embarrassingly parallel computation of individual trajectories.
We explored the capability of using the gradient estimator for efficient inference of the QGP properties by rediscovering the two transport coefficients $\hat{\kappa}$ and $\hat{\gamma}$ from synthetic data. To this end, we carried out large-scale simulations of the underlying Lindblad dynamics and performed gradient descent in the parameter space using an automated HPC workflow.
We find that, using the score-function gradient estimator, the optimizer converges smoothly to the true transport coefficient values, even with a comparatively small number of samples per step.
These results suggest that parameter inference in settings with expensive-to-evaluate Monte Carlo simulators can be approached with gradient descent, if an efficient gradient estimator can be formulated.

In the future, our method can be used to determine the transport coefficients of the quark-gluon plasma from experimental data for the nuclear modification factor, measured by CMS, ATLAS, and ALICE~\cite{ALICE:2020wwx,ATLAS:2022exb,CMS:2018zza,CMS:2023lfu}. The method will be especially useful, since refined master equations contain more than two transport coefficients~\cite{Brambilla:2024quarkonium}. In a higher-dimensional parameter space, gradient descent will be particularly valuable, since, e.g., Bayesian optimization becomes even more expensive.

Furthermore, it may be possible to conserve even more resources by designing a specialized optimizer. By incorporating the information of the gradient estimator, such an optimizer could choose to draw a smaller number of samples when preceding gradient magnitude is large realtive to its uncertainty, as it would still be guaranteed that the optimizer would move approximately in the right direction. Only when approaching the minimum, more accurate estimates of the gradient are required. In this way, depending on the loss landscape, a gradient-based optimization scheme could cope with a comparatively small number of total samples in the optimization run. Finally, the variance properties of the gradient estimator might be improved even further by leveraging advanced techniques from the reinforcement learning community~\cite{tucker2017rebar,grathwohl2017backpropagation,gu2015muprop}, which could further reduce the trajectory count required per optimization step.

%% file: Appendix.tex
\section{Lindblad equation for quarkonium in medium}
\label{app:quarksuppr}

We provide the explicit form of the Hamiltonian and the jump operators in Eq.~\eqref{eq:lindblad}, which are determined up to order $E/(\pi T)$ in Ref.~\cite{Brambilla:2022ynh}.
In three dimensions, there is a separate set of jump operators for each dimension, so Eq.~\eqref{eq:lindblad} becomes
\begin{align}
    \begin{split}
        \frac{d\rho(t)}{dt}=\mathcal{L}&\left[\rho(t)\right]=-i[H,\rho(t)]\\
    &+\sum_{n,i}\left[C_i^n\rho(t)C_i^{n\dagger}-\frac{1}{2}\left\{C_i^{n\dagger}C_i^n,\rho(t)\right\}\right],
        \label{eq:lindblad2}
    \end{split}
\end{align}
where $i=x,y,z$ enumerates the dimension.
The Hamiltonian for the bottomonium is given by
\begin{equation}
H=
\begin{pmatrix} 
h_s+ h_T & 0 \\
0 & h_o+\frac{N^2_c-2}{2(N^2_c-1)}h_T
\end{pmatrix},
\label{eq:HamiltonianQCDNLO}
\end{equation}
where the thermal effects on the Hamiltonian are contained in
\begin{equation}
    h_T = \frac{r^2}{2}\gamma +\frac{\kappa}{4MT}\{r_i,p_i\},
\end{equation}
where $r_i$ and $p_i$ denote the position and momentum operators respectively, and $r=|\vec{r}|$.
The $2\times 2$ matrix represents the singlet and octet color structure and the $h_{s(o)}$ denote the singlet and octet vacuum Hamiltonian respectively
\begin{equation}
    h_{s(o)}=\frac{p^2}{M}+V_{s(o)}(r),
\end{equation}
with $V_{s(o)}(r)$ the quarkonium potential from potential non-relativistic QCD. At leading order, we take these to be coulombic $V_s(r)=-\alpha/r$ and $V_o(r)=\alpha/(8r)$. Furthermore, $N_c$ denotes the number of colors, which for QCD will always be set to $N_c=3$.
The collapse operators $C^n_i$, on the other hand, are given by
\begin{align}
    C_i^0 &= \sqrt{\frac{\kappa}{N^2_c-1}}\begin{pmatrix}
        0 & K_i+\frac{\Delta V_{os}}{4T}r_i\\
        \sqrt{N^2_c-1}\left(K_i+\frac{\Delta V_{so}}{4T}r_i\right) & 0
    \end{pmatrix},\label{eq:QCDNLOC0}\\
    C_i^1 &= \sqrt{\frac{\kappa(N^2_c - 4)}{2(N^2_c-1)}}\begin{pmatrix}
        0 & 0\\
        0 & K_i\end{pmatrix},\label{eq:QCDNLOC1}
\end{align}
with 
\begin{equation}
    K_i = r_i+\frac{ip_i}{2MT},
\end{equation}
and $\Delta V_{uv} = V_u-V_v$. The transport coefficients $\kappa$ and $\gamma$ are defined in terms of the dimensionless quantities as $\kappa=\hat\kappa\,T^3$ and $\gamma = \hat\gamma\,T^3$.

\section{Quarkonium suppression Phenomenology}
\label{app:pheno}

To study the survival probability of the bottomonium throughout the evolution, we project the density matrix on quarkonium eigenstates $\ket{nl}$ defined as a solution of the Schrödinger equation
\begin{equation}
    \left(\frac{p^2}{M}+V_s(r)+\frac{l(l+1)}{Mr^2}\right)\ket{nl} = E\ket{nl}.
\end{equation}
The $1S$ bottomonium wave function $\ket{1S}$ would, e.g., correspond to the ground state of this Hamiltonian with $l=0$. We then interpret 
\begin{equation}
    p_{1S}(t) = \frac{\braket{1S|\rho(t)|1S}}{\braket{1S|\rho(0)|1S}},
    \label{eq:survival_prob}
\end{equation}
as the survival probability of the $1S$ bottomonium. The survival probabilities can be related to the nuclear modification factor $R_{AA}$, which measures the suppression of measured quarkonia states in heavy-ion collisions compared to proton-proton collisions. However, to do so, we have to take into account that, after surviving the QGP, excited quarkonium states might decay into lower lying states, effectively modifying the number of quarkonia measured in the detector. To account for this, an excited state feed-down is performed by introducing a feed-down matrix $F$~\cite{Brambilla:2020qwo}. The matrix $F$ acts on a vector of different quarkonium states. In general we will consider the experimental cross sections $\vec{\sigma}_\text{exp}$ as measured in $p-p$-collisions for the set of quarkonium states $\{\Upsilon(1S), \Upsilon(2S), \chi_{b0}(1P), \chi_{b1}(1P), \chi_{b2}(1P),\allowbreak\Upsilon(3S),\allowbreak\chi_{b0}(2P),\allowbreak\chi_{b1}(2P),\chi_{b2}(2P)\}$. The feed-down matrix $F$ allows us to relate the experimental cross section to a direct production cross section $\vec{\sigma}_\text{direct}$ by $\vec{\sigma}_\text{exp}=F\vec{\sigma}_\text{direct}$. The direct production cross section can be thought of as the cross section after the initial stages of the production. At the same time, the feed-down matrix accounts for the decay of higher states during the trajectory of the quarkonium into the detector. To obtain the nuclear modification factor, we further introduce the survival matrix $S$, which carries the survival probabilities $p_i$ on its diagonal. The nuclear modification factor for the state $i$ can then be written as
\begin{equation}
    R^i_{AA} = \frac{(F\cdot S\cdot\vec{\sigma}_\text{direct})^i}{\sigma^i_\text{exp}},
    \label{eq:RAA}
\end{equation}
where $\vec{\sigma}_\text{direct}$ is calculated as $\vec{\sigma}_\text{direct}=F^{-1}\vec{\sigma}_\text{exp}$. The numerator in this equation can be interpreted as taking the direct production of the different quarkonium states, then multiplying by the probability to survive the evolution throughout the QGP through $S$, and finally applying the feed-down of excited states through $F$.
By taking the ratio to the experimental cross section $\sigma^i_\text{exp}$, we compare the prediction with QGP effects against the measurement without QGP effects, effectively obtaining the suppression. We note that by calculating the direct production cross section from the experimental proton-proton cross-section, we neglect cold nuclear matter effects, which, however, are expected to be subleading in lead-lead collisions.

The properties of the QGP depend on the centrality of the collision, that is, how much the two nuclei overlap when colliding. Generally, a larger overlap results in higher temperatures and a plasma with a longer lifetime. The nuclear modification factor is therefore often reported in terms of the number of participating nucleons in the collision $N_\text{part}$, where a large number of participating nucleons corresponds to a significant overlap of the colliding nuclei. In practice, we use the temperature evolution of the plasma for different centralities obtained from hydrodynamics simulations and solve the Lindblad equation with different $T(t)$, assuming the temperature evolution to be quasistatic. These solutions lead to predictions for the survival probabilities and, ultimately, the nuclear modification factor in terms of the number of participating nucleons, which can be compared to experimental data.